\documentclass[prb,aps,showpacs,superscriptaddress,twocolumn]{revtex4-1}
\usepackage{latexsym}
\usepackage{amsfonts}
\usepackage{amssymb}
\usepackage{amsmath}
\usepackage{graphicx}

\begin{document}
\title{Mn-doping induced ferromagnetism and enhanced superconductivity in\\
$\rm Bi_{4-x}Mn_xO_4S_3$ ($0.075 \le x \le 0.15$)}
\date{\today}

\author{Zhenjie Feng$^{1,2,\rm a}$, Xunqing Yin$^1$, Yiming Cao$^1$, Xianglian Peng$^1$, Tian Gao$^3$, Chuan Yu$^1$, Jingzhe Chen$^1$, Baojuan Kang$^1$, Bo Lu$^4$, Juan Guo$^5$, Qing Li$^1$, Wei-Shiuan Tseng$^6$, Zhongquan Ma$^1$, Chao Jing$^1$, Shixun Cao$^{1,2,\rm a}$, Jincang Zhang$^{1,7}$ and N.-C. Yeh$^{6,\rm a}$}

\affiliation{Department of Physics, Shanghai University, Shanghai 200444, China\\
$^2$Shanghai Key Laboratory of High Temperature Superconductors, Shanghai 200444, China\\
$^3$Department of Physics, Shanghai University of Electric Power, Shanghai 201300, China\\
$^4$Laboratory for Microstructures, Shanghai University, Shanghai 200444, China\\
$^5$School of Physical Engineering, Zhengzhou University, Zhengzhou 450001, China\\
$^6$Department of Physics, California Institute of Technology, Pasadena, CA 91125, USA\\
$^7$Materials Genome Institute, Shanghai University, Shanghai 200444, China}

\begin{abstract}
We demonstrate that Mn-doping in the layered sulfides $\rm Bi_4O_4S_3$ leads to stable $\rm Bi_{4-x}Mn_xO_4S_3$ compounds that exhibit both long-range ferromagnetism and enhanced superconductivity for $\rm 0.075 \le x \le 0.15$, with a possible record superconducting transition temperature ($T_c$) $\sim 15$ K amongst all $\rm BiS_2$-based superconductors. We conjecture that the coexistence of superconductivity and ferromagnetism may be attributed to Mn-doping in the spacer $\rm Bi_2O_2$ layers away from the superconducting $\rm BiS_2$ layers, whereas the enhancement of $T_c$ may be due to excess electron transfer to $\rm BiS_2$ from the Mn$^{4+}$/Mn$^{3+}$-substitutions in $\rm Bi_2O_2$. This notion is empirically corroborated by the increased electron-carrier densities upon Mn doping, and by further studies of the $\rm Bi_{4-x}A_xO_4S_3$ compounds (A = Co, Ni; x = 0.1, 0.125), where the $T_c$ values remain comparable to that of the undoped $\rm Bi_4O_4S_3$ system ($ \sim 4.5$ K) due to lack of 4+ valences in either Co or Ni ions for excess electron transfer to the $\rm BiS_2$ layers. These findings therefore shed new light on feasible pathways to enhance the $T_c$ values of $\rm BiS_2$-based superconductors, although complete elucidation of the interplay between superconductivity and ferromagnetism in these  anisotropic layered compounds awaits the development of single crystalline materials for further investigation.\footnote{Nai-Chang Yeh (ncyeh@caltech.edu); Zhenjie Feng (fengzhenjie@shu.edu.cn); Shixun Cao (sxcao@shu.edu.cn). Correspondence should be addressed to N.-C.Y. and requests for materials should be addressed to either Z.-J. F. or N.-C. Y.}

\end{abstract}
\maketitle

\section{Introduction}

One of the commonalities among the cuprate and iron-based high-temperature superconductors is their layered structures.~\cite{BednorzJG1986,KamiharaY2008} Interestingly, even for conventional superconductors, the highest superconducting transition temperature ($T_c$) has been found in layered magnesium diboride $\rm MgB_2$.~\cite{NagamatsuJ2001} Recently, superconductivity with $T_c$ = 4.5 K was discovered in a new superconductor $\rm Bi_4O_4S_3$.~\cite{MizuguchiY2012} This compound has a layered structure composed of two superconducting $\rm BiS_2$ layers and spacer layers of $\rm Bi_4O_4(SO_4)_{1-x}$, where x indicates the deficiency of $\rm (SO_4)^{2-}$ ions at the interlayer sites. Since the discovery of $\rm Bi_4O_4S_3$, several other $\rm BiS_2$-based superconductors $\rm LnO_{1-x}F_xBiS_2$ (Ln = La, Ce, Pr, Nd) with the highest $T_c \sim$ 10.6 K have been reported.~\cite{MizuguchiY2012a,AwanaVPS2013,XingJ2012,JhaR2014,JhaR2013,DemuraS2013,Yazici2013} Both experimental and theoretical studies to date have indicated that the $\rm BiS_2$ layers play the role of the superconducting planes in these sulfide superconductors, similar to the $\rm CuO_2$ planes in the cuprate superconductors and the $\rm Fe_2An_2$ (An = P, As, Se, Te) layers in the iron-based superconductors.~\cite{BednorzJG1986,KamiharaY2008,YehNC2014} 

A major challenge facing this new class of layered superconductors is to optimize $T_c$ by exploring different spacer layers. Additionally, the effects of doping by either non-magnetic or magnetic elements are important issues for investigation. To date, suppression of superconductivity has been observed in the case of Cu and Ag substitutions for Bi in the $\rm Bi_4O_4S_3$ superconductor,~\cite{LiuY2013,TanSG2012} whereas coexistence of superconductivity and ferromagnetism has been reported in the $\rm CeO_{1-x}F_xBiS_2$~\cite{DemuraS2015,LeeJ2014} and $\rm Sr_{0.5}Ce_{0.5}FBiS_2$~\cite{LiL2015} systems at low temperatures. However, none of these doping effects are fully understood. 

Aiming at addressing the aforementioned issues, we report in this work our studies of 3$d$ transition-metal substitutions for Bi in $\rm Bi_4O_4S_3$by synthesizing $\rm Bi_{4-x}A_xO_4S_3$ (A = Mn, Co, Ni; $\rm 0.075 \le x \le 0.15$) compounds with conventional solid state reaction. We first focus on the investigation of $\rm Bi_{4-x}Mn_xO_4S_3$ because these results are most interesting and reveal a possible record $T_c \sim$ 15 K, and then perform comparative studies on $\rm Bi_{4-x}A_xO_4S_3$ (A = Co, Ni) in the Discussion section to elucidate the underlying physics. Based on our empirical findings, we suggest that the coexistence of superconductivity and long-range ferromagnetism in all $\rm Bi_{4-x}A_xO_4S_3$ (A = Mn, Co, Ni) compounds may be attributed to the selective doping of 3$d$ transition-metal elements in the spacer $\rm Bi_2O_2$ layers, whereas the enhancement of $T_c$ found only in Mn-doped samples may be due to substantial electron transfer from Mn$^{4+}$/Mn$^{3+}$-substitutions in $\rm Bi_2O_2$ to the superconducting $\rm BiS_2$ layers. 

\begin{figure*}
  \centering
  \includegraphics[width=6.8in]{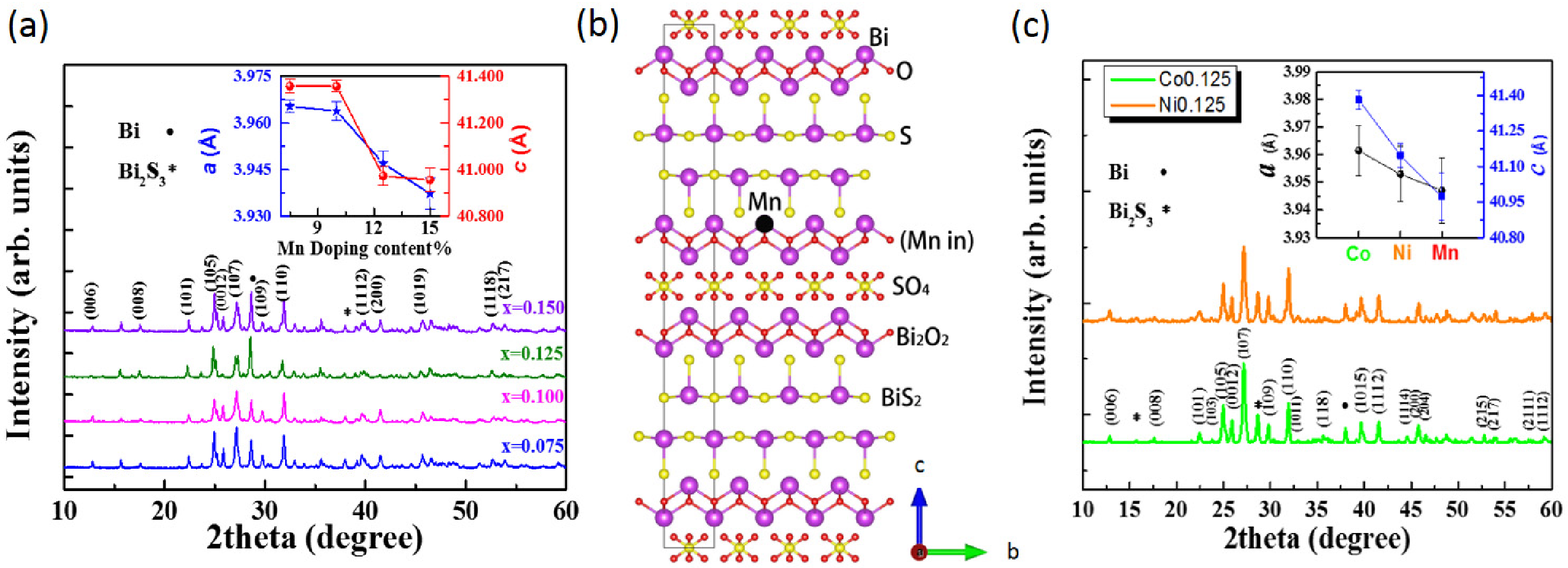}
\caption{(Color online) Structural properties of $\rm Bi_{4-x}A_xO_4S_3$ (A = Mn, Co, Ni) : (a) X-ray diffraction (XRD) patterns of $\rm Bi_{4-x}Mn_xO_4S_3$ ($\rm 0.075 \le x \le 0.15$). The inset shows the doping dependent variations of the in-plane and c-axis lattice parameters $a$ and $c$. (b) Schematics of the layered structure of $\rm Bi_{4-x}Mn_xO_4S_3$. (c) X-ray diffraction (XRD) spectral studies of $\rm Bi_{4-x}Co_xO_4S_3$ and $\rm Bi_{4-x}Ni_xO_4S_3$ for x = 0.125. The XRD spectra indicated that the lattice constants after 3$d$ transition-metal doping were all reduced relative to those of $\rm Bi_4O_4S_3$, and the values for different dopants followed the descending order of Co, Ni and Mn, as shown in the inset of (c).}
\label{Fig1}
\end{figure*}

\section{Experimental}

Bulk polycrystalline $\rm Bi_{4-x}Mn_xO_4S_3$ (x = 0.075, 0.1, 0.125, 0.15) and $\rm Bi_{4-x}A_xO_4S_3$ (A = Co, Ni; x = 0.1, 0.125) samples were synthesized by conventional solid state reaction method. For the $\rm Bi_{4-x}Mn_xO_4S_3$ samples, high purity Bi (99.99\%), $\rm Bi_2O_3$ (99.99\%), S (99.999\%), $\rm MnO_2$ (99.99\%) were first weighed in stoichiometric ratio and then grounded thoroughly in a glove box under high purity argon atmosphere. Next, the mixture was pressed into a pellet shape and sealed in an evacuated quartz tube (10$^{-4}$ Torr). The pellet was then heated up to 510 $^{\circ}$C and kept for 10 hours. After cooling the pellet to room temperature, the product was well mixed again by regrinding, pressed into a pellet shape, and then annealed at 510 $^{\circ}$C for another 10 hours. The samples thus obtained looked black and were hard. It is important to note that the sample may not be heated above 550 $^{\circ}$C. Otherwise S-O gas would be produced and could result in explosion of the quartz tube.~\cite{MizuguchiY2012} Similar procedures were applied to the synthesis of the $\rm Bi_{4-x}A_xO_4S_3$ (A = Co, Ni; x = 0.1, 0.125) samples.

The crystal structures of all samples were characterized by X-ray powder diffraction (XRD, 18 kW D/MAX 2550) using the Cu-$K_{\alpha}$ radiation. The lattice constants were calculated from the 2$\theta$ values and the Miller indices by using the Jade 6.5 software. After XRD studies, these polycrystalline samples were cut into rectangular shape and polished for electrical resistivity measurements. The electrical resistivity was measured with a standard four-terminal method covering temperature range from 3 to 300 K in a Physical Property Measurement System (PPMS-9, Quantum Design, Inc.). Typical current densities used for the resistive measurements were $\sim$ 100 $\rm A/m^2$. No apparent dependence on the current density was found up to $\sim$ 2000 $\rm A/m^2$, whereas resistive signals became difficult to resolve for current densities significantly smaller than 100 $\rm A/m^2$. The magnetization and specific heat measurements were conducted using the same PPMS with Vibrating Sample Magnetometer (VSM) and specific heat options. The zero-field-cooling (ZFC) and field-cooling (FC) of the magnetic susceptibility measurements of the samples were performed in the warming process. Additionally, the carrier densities of bulk polycrystalline samples were determined from their normal-state Hall coefficients at 300 K by means of the van der Pauw method and with the use of the Hall Effect Measurement System CVM200 made by the East Changing Company.

\begin{figure}
  \centering
  \includegraphics[width=3.4in]{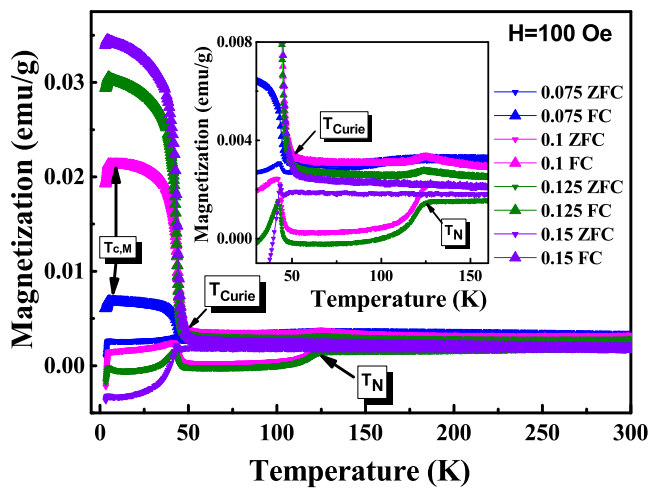}
\caption{(Color online) Temperature dependent magnetization of $\rm Bi_{4-x}Mn_xO_4S_3$ (x = 0.075, 0.10, 0.125, 0.15): Zero-field-cool (ZFC) and field-cool (FC) magnetization as a function of $T$ is shown in the main panel from 3 to 300 K with an external field $H$ = 100 Oe. For each magnetization curve, three characteristic temperatures are noteworthy: The N$\rm \acute{e}$el temperature ($T_N$) near $\sim 125$ K for x = 0.075, 0.10 and 0.125; the Curie temperature ($T_{\rm Curie}$) near $\sim$ 50 K; and the magnetization-determined superconducting transition temperature ($T_{c,M}$). The inset shows an expansion of the main panel over the temperature range where strong contrasts appear between the ZFC and FC curves.}
\label{Fig2}
\end{figure}

\section{Results and Analysis}

\subsection{Structural characterization}

The XRD patterns of different Mn-doped samples and the corresponding crystalline structure are shown in Fig.~\ref{Fig1}(a)-(b). These data suggest that all samples acquired the expected tetragonal phase (space group I4/mmm) with minor rhombohedra $\rm Bi_2S_3$ and Bi impurities, the latter being common occurrences in the $\rm Bi_4O_4S_3$ system~\cite{MizuguchiY2012,LiuY2013,TanSG2012,SrivastavaP2014} and exhibiting no superconductivity above 3 K.~\cite{MuntyanuF2006,WeitzelB1991,ChenB1997} The nominal Mn-doping (x) dependence of the lattice constants $a$ and $c$ are illustrated in the inset of Fig.~\ref{Fig1}(a). The general trend of decreasing lattice parameters with increasing Mn-doping is reasonable because the ionic radii of Mn are much smaller than that of Bi$^{3+}$.~\cite{Shannon1976} This trend is also indicative of successful incorporation of Mn-ions into the $\rm Bi_4O_4S_3$ unit cells. Similarly, XRD studies of $\rm Bi_{4-x}A_xO_4S_3$ (A = Co, Ni) also indicate that the lattice constants of $\rm Bi_{1-x}A_xO_4S_3$ were all reduced relative to those of $\rm Bi_4O_4S_3$, as shown in Fig.~\ref{Fig1}(c). Moreover, the lattice constants for different dopants followed the descending order of Co, Ni and Mn, as explicitly shown in the inset of Fig.~\ref{Fig1}(c).

\begin{figure*}
  \centering
  \includegraphics[width=5.4in]{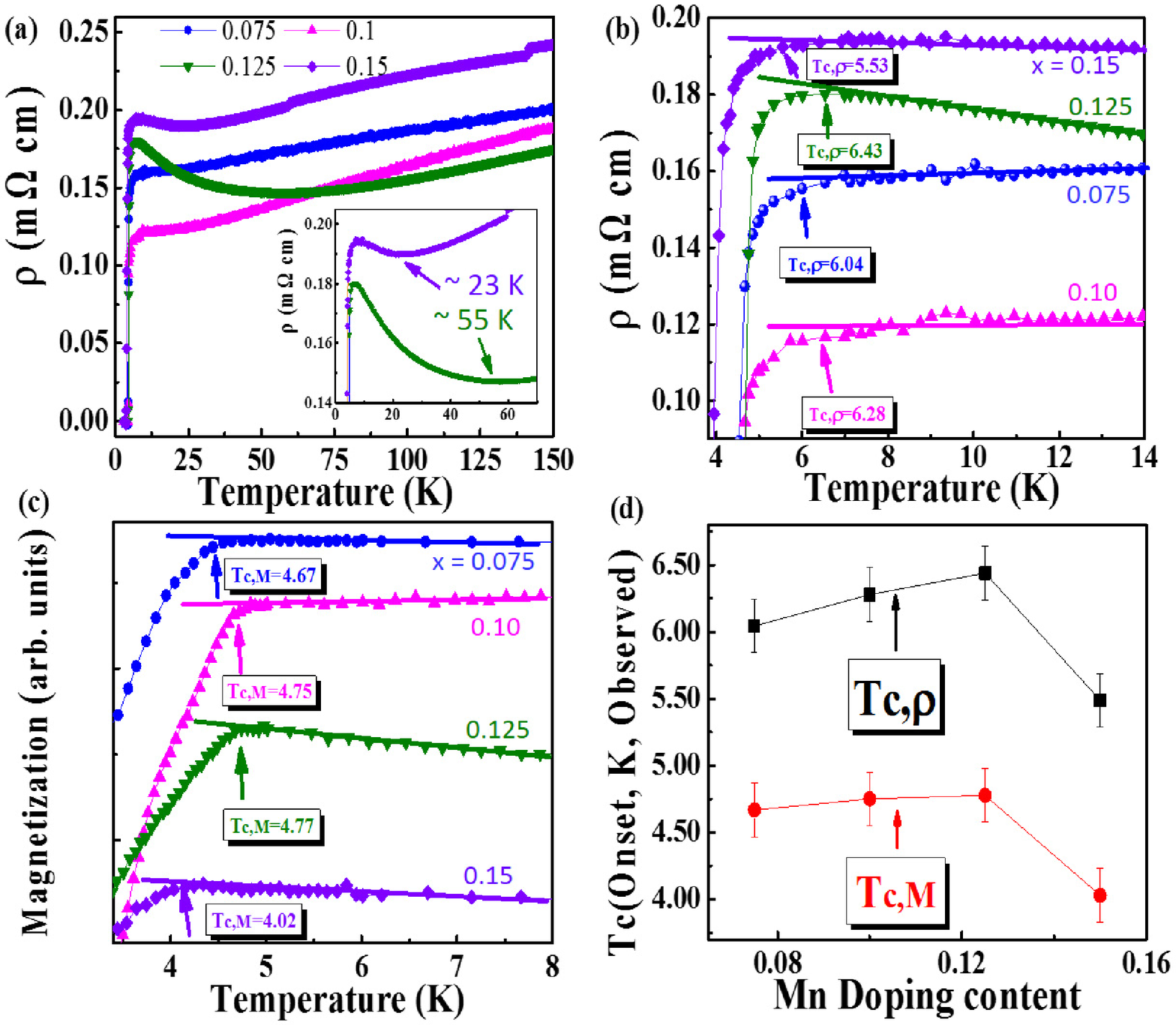}
\caption{(Color online) Characterization of the superconducting transition temperatures of $\rm Bi_{4-x}Mn_xO_4S_3$: (a) Resistivity ($\rho$) vs. temperature ($T$) behavior of $\rm Bi_{4-x}Mn_xO_4S_3$. The inset is the enlargement of the lower temperature regime for x = 0.125 and 0.15, showing Kondo-like resistive upturn at $T \sim 55$ K for x = 0.125 and at $T \sim 23$ K for x = 0.15. (b) Detailed $\rho$-vs.-$T$ curves of all samples near the onset of resistive superconducting transition ($T_{c,\rho}$), where $T_{c,\rho}$ (in units of K) exhibits slight decrease with increasing x. (c) ZFC-magnetization vs. $T$ behavior under $H$ = 100 Oe and near $T_{c,\rho}$ (in units of K). (d) Comparison of the Mn-doping level dependence of $T_{c,\rho}$ from resistive data and $T_{c,M}$ from magnetization measurements.}
\label{Fig3}
\end{figure*}

\subsection{$M$-vs.-$T$ and $\rho$-vs.-$T$ studies of $\rm Bi_{4-x}Mn_xO_4S_3$}

Temperature ($T$) dependent magnetization ($M$) of $\rm Bi_{4-x}Mn_xO_4S_3$ with x = 0.075, 0.10, 0.125 and 0.15 was studied under both zero-field-cool (ZFC) and field-cool (FC) conditions from 3 to 300 K and with an external field $H$ = 100 Oe, as illustrated in Fig.~\ref{Fig2}. For each magnetization curve, three characteristic temperatures are noteworthy: The N$\rm \acute{e}$el temperature ($T_N$) near $\sim 125$ K for x = 0.125 and 0.15, below which $M$ decreased due to the onset of antiferromagnetism; the Curie temperature ($T_{\rm Curie}$) near $\sim 50$ K for all samples, below which a rapid upturn followed by saturation in the FC magnetization curves appeared, suggesting the formation of long range ferromagnetism; and the temperature $T_{c,M} \sim 4.5$ K below which rapid decrease in magnetization occurred as the result of supercurrent-induced diamagnetism. Additionally, we note the dramatic contrasts between the ZFC and FC magnetization curves for $T_{c,M} < T < T_{\rm Curie}$ in all samples: The ZFC curves all exhibited an initial upturn of magnetization, signaling the onset of ferromagnetism, which was followed by gradual decrease and then a sharp downturn in magnetization. Interestingly, both the diamagnetic contribution in the ZFC curve and the magnitude of ferromagnetism in the FC curve increased with increasing x.   

To better understand the interplay of magnetism and superconductivity, we conducted measurements of resistivity ($\rho$) vs. $T$ on $\rm Bi_{4-x}Mn_xO_4S_3$. As shown in Fig.~\ref{Fig3}(a), all samples reached zero resistance at low temperatures. On the other hand, the resistivity of samples with lower Mn-doping levels (x = 0.075, 0.10) exhibited monotonic temperature dependence up to 150 K, whereas a resistive upturn~\cite{AndreiN1983} appeared at $\sim 55$ K and $\sim 23$ K for higher Mn-doping levels x = 0.125 and 0.15, respectively. Given the highly anisotropic, layered nature of these $\rm BiS_2$-based compounds, the physical origin for this doping dependent resistive upturn cannot be fully uncovered without the availability of single crystalline materials. Nonetheless, a feasible mechanism that contributes to the resistive upturn is the occurrence of Kondo resonance at $T < T_K$, where $T_K$ denotes the Kondo temperature. In this scenario, a lower $T_K$ for a sample with a higher Mn-doping level would be consistent with stronger ferromagnetism and a sharper Kondo resonance of a linewidth $\sim T_K$.~\cite{AndreiN1983} Moreover, the formation of Kondo clouds below $T_K$ could help screen localized magnetic moments and so would be important to the appearance of singlet superconductivity at $T_c < T_K$.~\cite{PfleidererC2009} However, the onset temperature $T_{c,\rho}$ for rapid decrease in resistivity did not exhibit strong doping dependence (Fig.~\ref{Fig3}(a)-(b)), and the $T_{c,\rho}$ values for all doping levels were generally higher than $T_{c,M}$, the onset temperature for rapid ZFC and FC magnetization decrease, although both $T_{c,\rho}$ and $T_{c,M}$ followed a similar non-monotonic trend (Fig.~\ref{Fig3}(c)-(d)). Given the complex conduction paths and magnetic domain structures in typical polycrystalline samples, the doping dependent resistive upturn above $T_c$ may be in part but cannot be entirely attributed to the occurrence of Kondo resonance. 

\begin{figure*}
  \centering
  \includegraphics[width=5.4in]{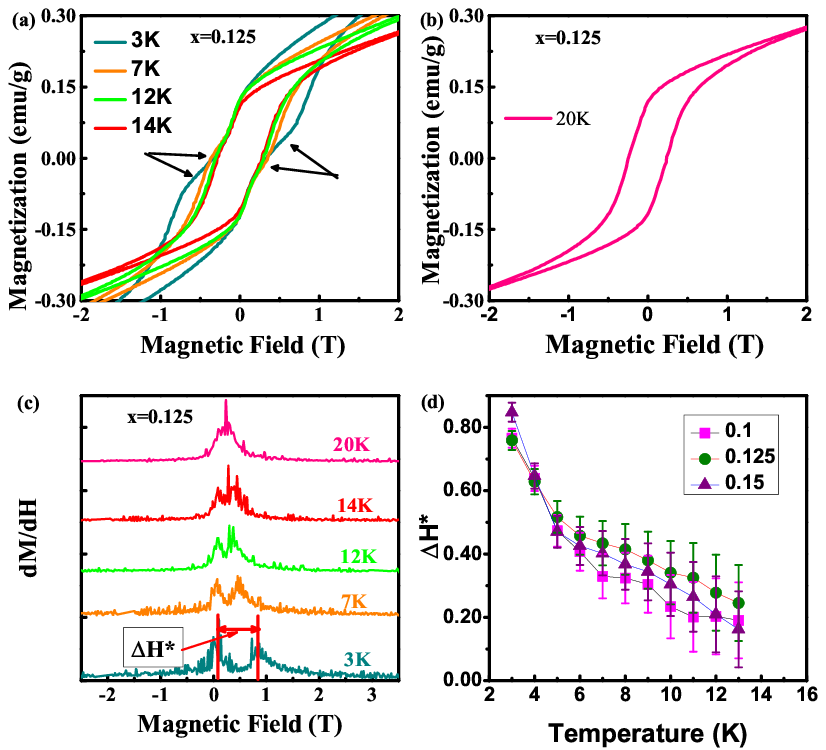}
\caption{(Color online) Unraveling the hidden superconducting transition from hysteretic $M$-vs.-$H$ curves: (a) Magnetization ($M$) vs. magnetic field ($H$) at $T$ = 3, 7, 12 and 14 K for x = 0.125, showing apparent anomalous features (indicated by the black arrows) associated with each magnetic hysteresis loop. These features are more pronounced at low $T$, and diminish with increasing $T$. (b) $M$-vs.-$H$ curve for $T$ = 20 K, which is consistent with a standard hysteresis loop for ferromagnets. (c) The ascending branches of ($dM/dH$)-vs.-$H$ curves at $T$ = 3, 7, 12, 14 and 20 K, respectively for x = 0.125, showing double-peak features for $T \le 14$ K and a single-peak feature at $T$ = 20 K. The magnetic field difference between the double-peak features of each ($dM/dH$)-vs.-$H$ curve is defined as $\Delta H^{\ast}$. (d) $\Delta H^{\ast}$-vs.-$T$ data for x = 0.10, 0.125 and 0.15. The $\Delta H^{\ast}$ values for all samples were found to approach zero at $T = (16 \pm 2)$ K $\sim T_c$. The nearly doping independent $T_c$ may result from competing effects of increasing carrier densities and ferromagnetism.}
\label{Fig4}
\end{figure*}

Generally speaking, the $T_{c,M}$ values determined from the onset of rapid FC magnetization decrease could not be representative of the intrinsic superconducting transition, because the coexistence of ferromagnetism and superconductivity would obscure the onset of the Meissner effect. Similarly, the polycrystalline nature of our $\rm Bi_{4-x}Mn_xO_4S_3$ samples could significantly reduce the $T_{c,\rho}$ values below the intrinsic superconducting transition temperature $T_c$ because of the inter-granular weak-link effects.~\cite{DurrellJ2011} Hence, additional thermodynamic measurements of $M$-vs.-$H$ at $T < T_{\rm Curie}$ and specific heat ($C$)-vs.-$T$ studies in both zero and finite magnetic fields were necessary to unravel the true $T_c$ values of $\rm Bi_{4-x}Mn_xO_4S_3$. 

\begin{figure*}
  \centering
  \includegraphics[width=5.4in]{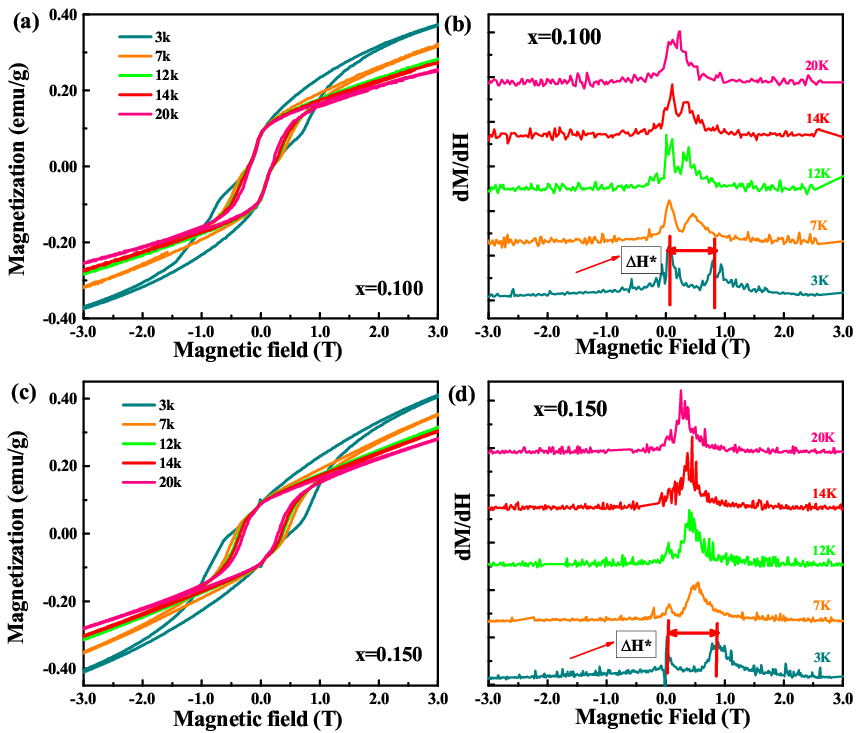}
\caption{(Color online) Isothermal $M$-vs.-$H$ loops of $\rm Bi_{4-x}Mn_xO_4S_3$ with x = 0.10 and 0.15: (a) The hysteresis loops for $\rm Bi_{3.9}Mn_{0.1}O_4S_3$ at $T$ = 3, 7, 12, 14 and 20 K. (b) The $dM/dH$-vs.-$H$ data for $\rm Bi_{3.9}Mn_{0.1}O_4S_3$ at $T$ = 3, 7, 12, 14 and 20 K, showing decreasing $\Delta H^{\ast}$ with increasing $T$. (c) The hysteresis loops for $\rm Bi_{3.85}Mn_{0.15}O_4S_3$ at $T$ = 3, 7, 12, 14 and 20 K. (d) The $dM/dH$-vs.-$H$ data for $\rm Bi_{3.85}Mn_{0.15}O_4S_3$ at $T$ = 3, 7, 12, 14 and 20 K, showing decreasing $\Delta H^{\ast}$ with increasing $T$.}
\label{Fig5}
\end{figure*}

\subsection{$M$-vs.-$H$ studies of $\rm Bi_{4-x}A_xO_4S_3$}

In Fig.~\ref{Fig4}(a) and (b) we show the hysteretic $M$-vs.-$H$ loops for x = 0.125 at {\it low} and {\it high} temperatures, respectively. Specifically, the {\it low}-temperature behavior in Fig.~\ref{Fig4}(a) with $T$ = 3, 7, 12 and 14 K refers to the appearance of anomalous features associated with each magnetic hysteresis loop. These features diminished with increasing $T$. In contrast, the {\it high}-temperature behavior as manifested in Fig.~\ref{Fig4}(b) for the $M$-vs.-$H$ loop at $T$ = 20 K reveals a standard magnetic hysteresis loop for a ferromagnetic material. We attribute the difference between the low- and high-temperature behaviors to the onset of superconductivity in the former. Specifically, we consider a standard although much smaller superconducting magnetization loop~\cite{YeshurunY1996} superposed on top of the ferromagnetic hysteretic loop. Both the isothermal ascending and descending branches of the $M$-vs.-$H$ loop at $T < T_c$ would deviate from the typical ferromagnetic hysteresis loop due to the presence of supercurrents. Hence, by considering the derivative $dM/dH$ of either the ascending or descending branch of the $M$-vs.-$H$ curve at a constant $T$, we expect one peak associated with the inflection point of a standard ferromagnetic $M$-vs.-$H$ curve at $T_c < T < T_{\rm Curie}$. In contrast, an additional peak in the $dM/dH$-vs.-$H$ curve is expected near $H$ = 0 for $T < T_c$ because of the appearance of supercurrents,~\cite{YeshurunY1996} which is indeed confirmed by the data shown in Fig.~\ref{Fig4}(c). 

We may define the magnetic field difference between the two peaks in $dM/dH$ as $\Delta H^{\ast} (T)$, which is a measure of the supercurrent.~\cite{YeshurunY1996} Therefore, we expect $\Delta H^{\ast} (T)$ to decrease with increasing $T$ and vanish at $T > \sim T_c$, which is consistent with the empirical finding shown in Fig.~\ref{Fig4}(d), where $\Delta H^{\ast} (T)$ approaches 0 at $T = (16 \pm 2)$ K for x = 0.125. Similar behavior has also been confirmed for x = 0.10 and 0.15, as shown in Fig.~\ref{Fig5}(a)-(d). These results suggest that $T_c \sim (16 \pm 2)$ K for x = 0.10, 0.125 and 0.15, whereas the $\Delta H^{\ast}$ values from the $dM/dH$ curves for x = 0.075 have been difficult to determine due to significantly smaller $M$-vs.-$H$ loops. 

\begin{figure}
  \centering
  \includegraphics[width=3.2in]{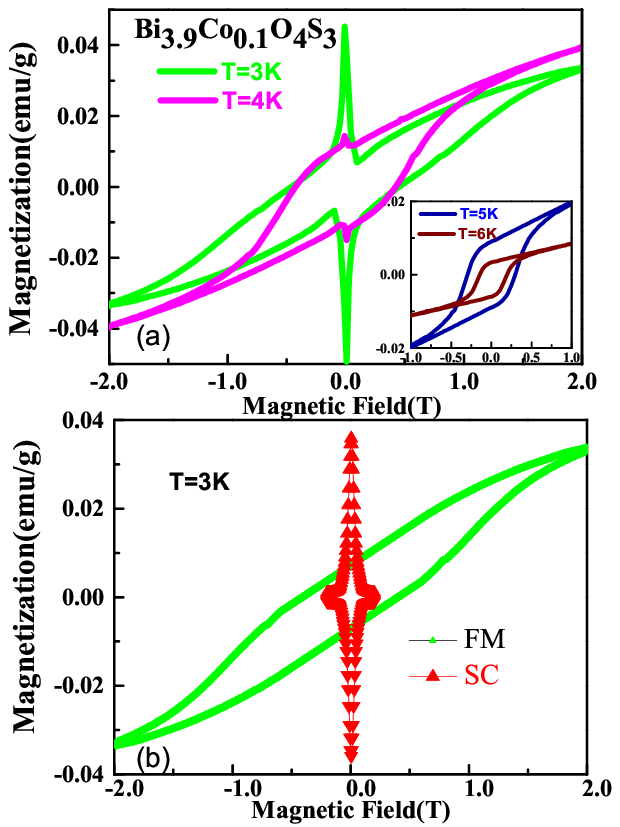}
\caption{(Color online) Isothermal $M$-vs.-$H$ loops of $\rm Bi_{4-x}Co_xO_4S_3$ with x = 0.125: (a) The hysteretic loops at $T$ = 3 and 4 K (main panel), and $T$ = 5 and 6 K (inset). The hysteric loops in the main panel reveal sharp anomalies, whereas those in the inset are consistent with typical ferromagnetic hysteretic loops. (b) The $M$-vs.-$H$ loop at $T = 3$ K in (a) can be decomposed into a ferromagnetic (green) and a superconducting (red) hysteretic loops from theoretical simulations.}
\label{Fig6}
\end{figure}

Our attribution of the anomalies in $M$-vs.-$H$ loops of $\rm Bi_{4-x}Mn_xO_4S_3$ to superconductivity can be further corroborated by studies of the $M$-vs.-$H$ loops of a related system $\rm Bi_{4-x}Co_xO_4S_3$ with x = 0.125. As we shall elaborate further in the Discussion section, the $\rm Bi_{3.875}Co_{0.125}O_4S_3$ compound is also a superconductor with $T_c \sim 4.8$ K and long-range ferromagnetism at low temperatures. Therefore, the $M$-vs.-$H$ loops of $\rm Bi_{3.875}Co_{0.125}O_4S_3$ also revealed anomalies induced by superconducting persistent currents at $T < T_c$, as shown in Fig.~\ref{Fig6}(a). On the other hand, the ferromagnetism in $\rm Bi_{3.875}Co_{0.125}O_4S_3$ was much weaker than that of $\rm Bi_{4-x}Mn_xO_4S_3$ (to be detailed in the Discussion section) so that the superconducting contributions in the $M$-vs.-$H$ loops could be much better revealed. As exemplified in Fig.~\ref{Fig6}(b), the $M$-vs.-$H$ loop of $\rm Bi_{3.875}Co_{0.125}O_4S_3$ taken at $T = 3$ K $< T_c \sim 4.8$ K could be decomposed into the superposition of a ferromagnetic loop and a superconducting loop. Moreover, we note that the superconducting contribution thus derived from the $M$-vs.-$H$ data of $\rm Bi_{4-x}Co_xO_4S_3$ was in good agreement with the total $M$-vs.-$H$ loop found in the non-magnetic, superconducting parent compound $\rm Bi_4O_4S_3$.~\cite{SinghSK2012} These results therefore reaffirmed our notion that the hysteretic $M$-vs.-$H$ loops of $\rm Bi_{4-x}A_xO_4S_3$ (A = Mn, Co, Ni) consisted of contributions from both superconductivity and ferromagnetism. 

\begin{figure}
  \centering
  \includegraphics[width=3.4in]{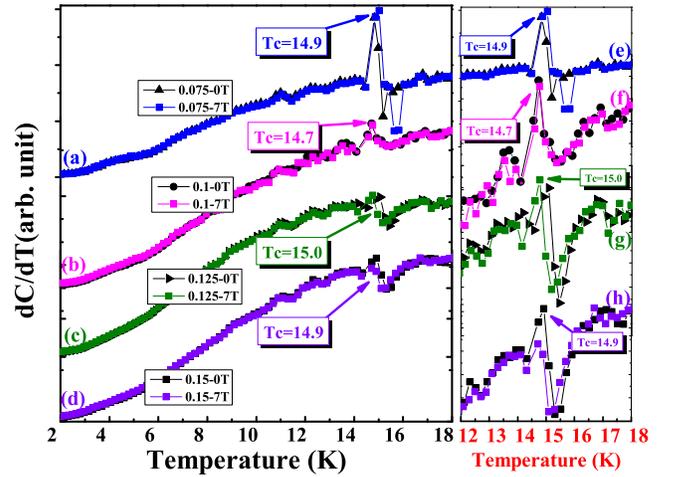}
\caption{(Color online) Specific heat ($C$) measurements as a function of $T$: (a)-(d) Temperature derivative of the specific heat ($dC/dT$)-vs.-$T$ curves for x = 0.075, 0.10, 0.125 and 0.15, respectively. The measurements were carried out for $H$ = 0 (black) and $H$ = 7 T (red). For each doping level, a sharp feature appears near $T \sim 15$ K. Taking the temperature where $dC/dT$ reaches maximum as $T_c$, we identify the $T_c$ value (in units of K) for each doping level from (a)-(d) and (e)-(h). The $T_c$ values ($\sim 15$ K) were comparable within experimental error for all samples in $H = 0$, similar to the findings shown in Fig.~\ref{Fig4}. On the other hand, slight downshifts ($\sim 0.3$ K) can be found for all samples under $H$ = 7 T, which is suggestive of a large upper critical field $H_{c2}$, as discussed in the text.}
\label{Fig7}
\end{figure}

\subsection{$C$-vs.-$T$ studies of $\rm Bi_{4-x}Mn_xO_4S_3$}

To further verify the intrinsic $T_c$ values of $\rm Bi_{4-x}Mn_xO_4S_3$, we conducted specific heat ($C$) studies on these samples. The measurements were carried out at both $H$ = 0 (black) and $H$ = 7 T (red), and the resulting ($dC/dT$)-vs.-$T$ curves for all doping levels x = 0.075, 0.10, 0.125 and 0.15 are respectively illustrated in Fig.~\ref{Fig7}~(a)-(d) and further detailed in Fig.~\ref{Fig7}(e)-(h). We find that a sharp and asymmetric feature appeared near $T \sim 15$ K for all curves taken at $H = 0$. If we attribute the temperature where maximum $dC/dT$ appeared to $T_c$, the $T_c$ values were found to be nearly doping independent ($T_c \sim 15$ K), consistent with the results obtained from the $dM/dH$ studies in Figs.~\ref{Fig4} and ~\ref{Fig5}. This finding of $T_c \sim 15$ K is the highest superconducting transition temperature reported to date among the $\rm BiS_2$-based superconductors. 

We further note that the peak position of each ($dC/dT$) curve taken at $H = 7$ T exhibited a small ($0.3 \sim 0.5$ K) downshift relative to that taken at $H = 0$ (Fig.~\ref{Fig7}~(e)-(h)), and the peak height also diminished with magnetic field. The small downshift of the peak position at $H = 7$ T is suggestive of a relatively strong upper critical field $H_{c2}(0)$. Using the Werthamer-Helfand-Hohenberg (WHH) theory~\cite{WHH1966} for the upper critical field, we find that the formula $H_{c2}(0) = - 0.69 T_c \lbrack dH_{c2} (T) / dT \rbrack _{T_c}$ for $T \to T_c ^-$ yields $H_{c2}(0)$ values in the range of $145 \sim 240$ T if we take $T_c = 15$ K. 

Although the errors for these $H_{c2}(0)$ estimates are likely significant because of our limited specific heat data, we note that the $H_{c2}(0)$ value of the parent compound $\rm Bi_4O_4S_3$ with $T_c = 4.5$ K was $\sim 21$ T,~\cite{MizuguchiY2012} suggesting that a relatively large $H_{c2}(0)$ value (on the order of $\sim 10^2$ T) for $T_c \sim 15$ K could be reasonable if the slope $|dH_{c2} (T) / dT|_{T_c}$ of $\rm Bi_{4-x}Mn_xO_4S_3$ was comparable to or even larger than that of the parent compound $\rm Bi_4O_4S_3$. On the other hand, in contrast to the studies of non-magnetic $\rm Bi_4O_4S_3$,~\cite{MizuguchiY2012} the $H_{c2}(0)$ values of magnetic $\rm Bi_{4-x}Mn_xO_4S_3$ could not be directly derived from the standard field-dependent resistive measurements due to the absence of $H$-dependent resistive transitions near $T_c \sim 15$ K and the fact that the WHH theory is only applicable to studies near $T_c$. Ultimately, better determination of the $H_{c2}(0)$ values of $\rm Bi_{4-x}Mn_xO_4S_3$ will rely on future availability of single crystalline materials, particularly given the highly anisotropic nature of these layered superconductors. 

\begin{figure*}
  \centering
  \includegraphics[width=5.4in]{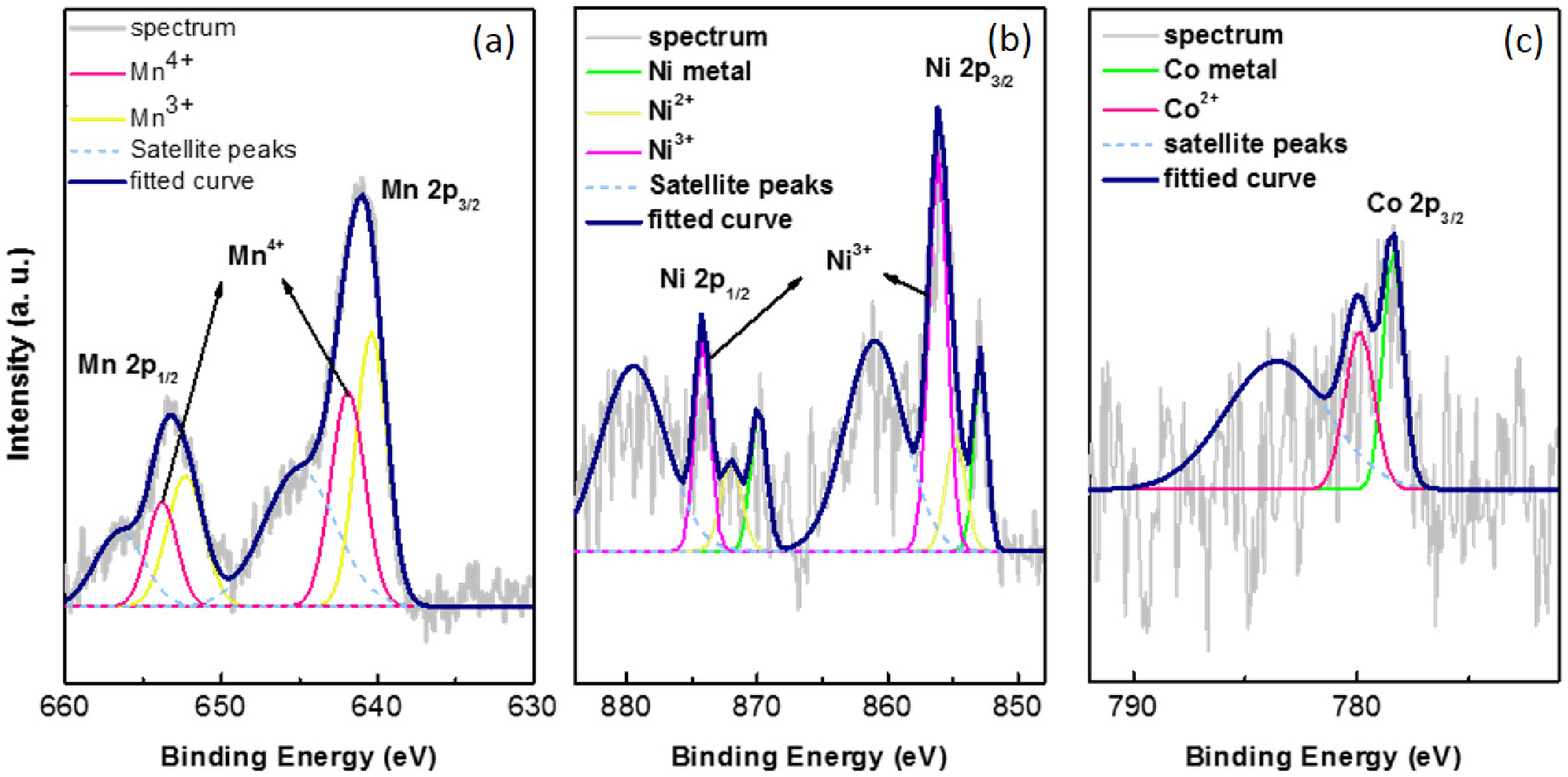}
\caption{(Color online) X-ray photoelectron spectroscopy (XPS) of $\rm Bi_{4-x}A_xO_4S_3$ with A = Mn, Co, Ni and x = 0.125: (a) Analysis of the XPS data for the Mn-$2p_{3/2}$ and Mn-$2p_{1/2}$ binding energies of $\rm Bi_{3.875}Mn_{0.125}O_4S_3$, showing that the valences of Mn-ions were mixtures of Mn$^{4+}$ (primary) and Mn$^{3+}$ (secondary),~\cite{NesbittHW1998} which gave rise to excess electron transfer to the $\rm BiS_2$ layers. Analysis of the XPS data of $\rm Bi_{3.875}Ni_{0.125}O_4S_3$, showing the valence of Ni being Ni$^{3+}$ (primary), metallic Ni and Ni$^{2+}$ (secondary).~\cite{ChenYS2012} The presence of minor metallic peaks suggested that a small fraction of Ni was not fully doped into $\rm Bi_4O_4S_3$. (c) Analysis of the XPS data of $\rm Bi_{3.875}Co_{0.125}O_4S_3$, showing the valences of Co-ions being metallic Co (primary) and Co$^{2+}$ (secondary),~\cite{DominguezM2010} suggesting that Co was not fully doped into $\rm Bi_4O_4S_3$, similar to the situation encountered in Ni-doped samples.}
\label{Fig8}
\end{figure*}

\section{Discussion}

\subsection{Possible physical origin for enhanced superconductivity and coexisting ferromagnetism in $\rm Bi_{4-x}Mn_xO_4S_3$}

The observation of both enhanced superconductivity and long-range ferromagnetism in the Mn-doped $\rm Bi_4O_4S_3$ samples may be associated with the unique location and valences of the Mn-ions in the $\rm Bi_4O_4S_3$ compound. First, Mn-ions may be preferentially located in the $\rm Bi_2O_2$ spacers rather than in the $\rm BiS_2$ layers so that severe lattice distortion from the large size differences between the Bi- and Mn-ions can be prevented in the superconducting $\rm BiS_2$ layers. Additionally, XPS studies on $\rm Bi_{4-x}Mn_xO_4S_3$ revealed that the valences of Mn-ions are found to be 3+ and 4+ (Fig.~\ref{Fig8}(a)). The substitution of Mn$^{4+}$ for Bi$^{3+}$ would result in excess electron doping and contribute to the electronic density of states at the Fermi level in the electron-type $\rm Bi_4O_4S_3$ superconducting system, thus enhancing $T_c$. 

The aforementioned conjecture of increased electronic carrier densities from Mn-doping to the $\rm Bi_4O_4S_3$ compound has indeed been corroborated by our Hall effect measurements of both pure $\rm Bi_4O_4S_3$ and $\rm Bi_{4-x}Mn_xO_4S_3$ samples. Specifically, the Hall effect measurements were carried out at 300 K with an applied magnetic field $H = 0.483$ T and an applied current $I = 10$ mA, and the Hall resistivity for each sample was obtained by averaging readings from multiple measurements on different contact positions of the sample via the van der Pauw method. For the parent compound $\rm Bi_4O_4S_3$, the normal-state Hall coefficient $R_H$ was found to be $(1.302 \pm 0.005) \times 10^{-4}$ $\rm m^3 C^{-1}$, which corresponded to a bulk carrier density of $n = (6.056 \pm 0.028) \times 10^{22}$ $\rm m^{-3}$. In contrast, for Mn-doped samples $\rm Bi_{4-x}Mn_xO_4S_3$ with x = 0.125, the normal-state Hall coefficient was found to be $R_H = (3.716 \pm 0.243) \times 10^{-5}$ $\rm m^3 C^{-1}$, which yielded a bulk carrier density $n = (1.688 \pm 0.120) \times 10^{23}$ $\rm m^{-3}$, more than 2.5 times that of $\rm Bi_4O_4S_3$.  

It is also interesting to note that the correlation between enhanced superconductivity and increased electron carrier densities in the $\rm BiS_2$-based superconductors has also been observed in the $\rm La_{1-x}M_xOBiS_2$ (M = Th, Hf, Zr, Ti) system,~\cite{Yazici2013a} where substitutions of tetravalent Th$^{4+}$, Hf$^{4+}$, Zr$^{4+}$ and Ti$^{4+}$ ions for trivalent La$^{3+}$ could induce superconductivity with $T_c$ up to 2.85 K while substitutions of divalent Sr$^{2+}$ for La$^{3+}$ could not yield superconductivity.
 
In addition to the effect of contributing excess carrier densities in the superconducting $\rm BiS_2$ layers, magnetic Mn-ions could give rise to long-range ferromagnetism in $\rm Bi_{4-x}Mn_xO_4S_3$ without directly affecting the Cooper pairing within the $\rm BiS_2$ layers at $T \ll T_K$ if they were confined within the $\rm Bi_2O_2$ spacer layers and coupled via the RKKY interaction.~\cite{RudermanMA1954,KasuyaT1956,YosidaK1957} Furthermore, the significant energy separation between the localized 3$d$ orbitals responsible for magnetism and the Fermi level within the itinerant 6$p$ orbitals for superconductivity also provides a favorable condition for coexisting superconductivity and ferromagnetism in $\rm Bi_{4-x}Mn_xO_4S_3$. In this context, we speculate that the observation of coexisting ferromagnetism and superconductivity in $\rm CeO_{1-x}F_xBiS_2$ and $\rm Sr_{0.5}Ce_{0.5}FBiS_2$ compounds~\cite{DemuraS2015,LeeJ2014,LiL2015} may also be attributed to the confinement of magnetic moments in the $\rm CeO_{1-x}F_x$ and $\rm Sr_{1-x}Ce_xF$ spacer layers. 

In principle, the location of Mn ions in $\rm Bi_{4-x}Mn_xO_4S_3$ may be identified by applying the Rietveld method to analyze the x-ray diffraction spectra.~\cite{RietveldHM1969} However, our attempts of using the Rietveld refinement could not conclude that Mn ions only substituted Bi ions in the $\rm Bi_2O_2$ layers because of too many fitting parameters. Future x-ray or neutron scattering experiments on single crystalline materials would be the best approach to conclusively determine the position of the doped Mn-ions.

Our conjecture of Cooper-pairing preservation in the $\rm BiS_2$ layers from the influence of adjacent ferromagnetic spacers is analogous to the findings in cuprate superconductivity, where substitutions of strong magnetic moments (such as Gd, Eu and Sm) in layers other than the $\rm CuO_2$ planes do not result in noticeable degradation of superconductivity.~\cite{ZapfV2005,PaulDM1988,GoldmanA1987} On the other hand, the significantly suppressed $T_{c,\rho}$ relative to $T_c$ in $\rm Bi_{4-x}Mn_xO_4S_3$ is likely due to the weak-link nature and ferromagnetic domain-induced superconducting phase fluctuations in these polycrystalline samples. 

Finally, we note that the large $H_{c2}(0)$ value estimated from our specific heat data also implies that the $\rm Bi_{4-x}Mn_xO_4S_3$ system can support superconductivity under substantially large effective magnetic fields, whether the fields are from external sources or due to local magnetic moments. The coexistence of superconductivity and ferromagnetism could also result in stronger superconducting fluctuations and contribute to a larger upper critical field. This interplay of superconductivity and ferromagnetism on the magnitude of the upper critical field of $\rm Bi_{4-x}Mn_xO_4S_3$ is an interesting issue worthy of further theoretical investigation.  

\subsection{Comparative studies of $\rm Bi_{4-x}Co_xO_4S_3$ and $\rm Bi_{4-x}Ni_xO_4S_3$}

\begin{figure*}
  \centering
  \includegraphics[width=5.4in]{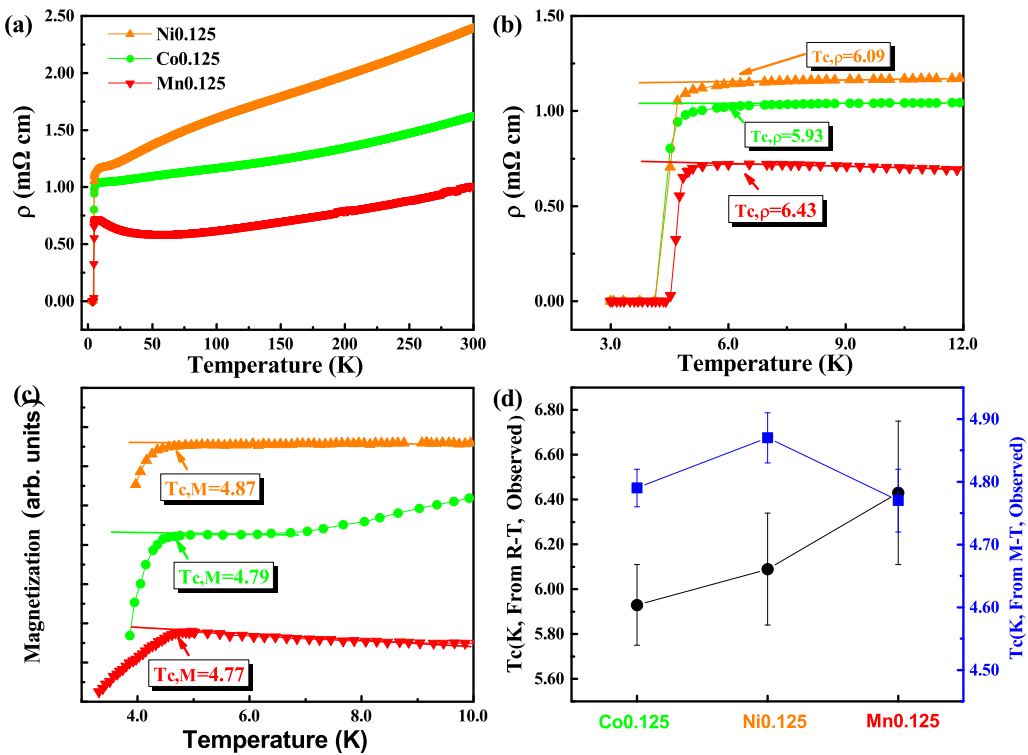}
\caption{(Color online) Resistive and magnetic characterizations of $\rm Bi_{4-x}A_xO_4S_3$ (A = Co, Ni, Mn and x = 0.125): (a) Comparison of the $\rho$-vs.-$T$ data up to $T = 150$ K for different 3$d$ transition-metal substitutions in the $\rm Bi_4O_4S_3$ system. (b) Comparison of the $\rho$-vs.-$T$  data at low temperatures (up to $T = 7.5$ K) for different 3$d$ transition-metal substitutions. (c) Comparison of the low-temperature (up to $T = 7.5$ K) ZFC $M$-vs.-$T$ data for different 3$d$ transition-metal substitutions. (d) Comparison of the $T_{c,\rho}$ and $T_{c,M}$ values for different 3$d$ transition-metal substitutions in $\rm Bi_4O_4S_3$. We note that the $T_{c,\rho}$ values exhibit anti-correlation with the lattice constants shown in the inset of Fig.~\ref{Fig1}(c).}
\label{Fig9}
\end{figure*}

\begin{figure}
  \centering
  \includegraphics[width=3.2in]{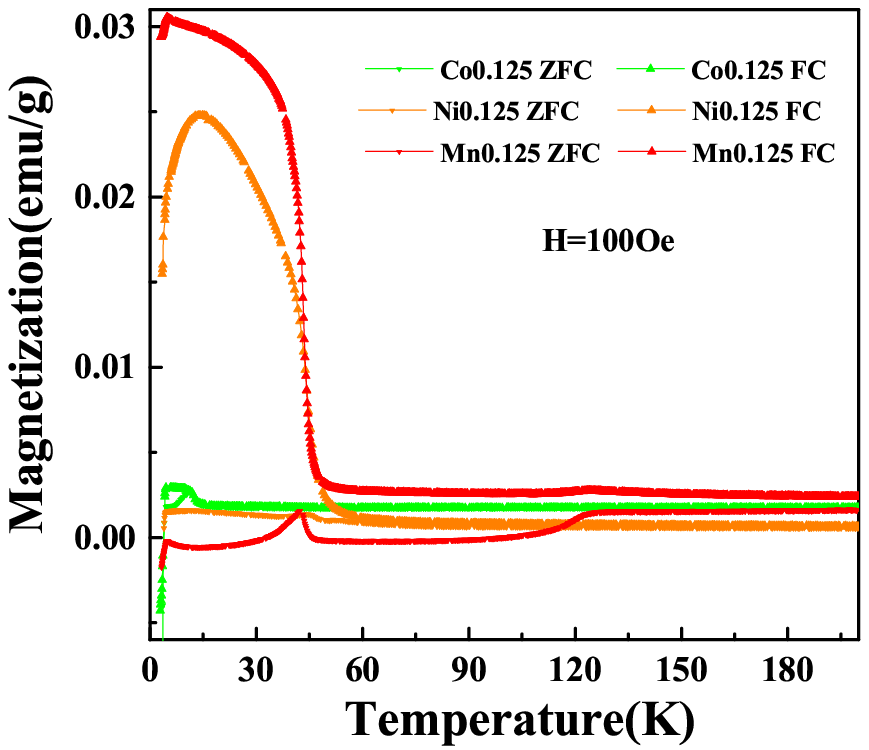}
\caption{(Color online) ZFC and FC $M$-vs.-$T$ data of $\rm Bi_{4-x}A_xO_4S_3$ (A = Co, Ni, Mn and x = 0.125): The ZFC $M$-vs.-$T$ curves for samples doped with Co and Ni always revealed positive magnetization except when $T < 4.5$ K, whereas their FC $M$-vs.-$T$ curves all exhibited strong increase in magnetization, suggesting the presence of long-range ferromagnetism. In contrast, only the ZFC $M$-vs.-$T$ curve for the Mn-doped sample $\rm Bi_{3.875}Mn_{0.125}O_4S_3$ exhibited diamagnetism for $T < 40$ K whereas the FC $M$-vs.-$T$ curve also revealed strong enhancement of positive magnetization below 50 K, suggesting the appearance of long range ferromagnetism.}
\label{Fig10}
\end{figure}

We have also investigated the effects of other 3$d$ transition-metal doping on $\rm Bi_4O_4S_3$, and found that ferromagnetism also coexists with superconductivity for $\rm Bi_{4-x}Co_xO_4S_3$ and $\rm Bi_{4-x}Ni_xO_4S_3$ with x = 0.1 and 0.125. As exemplified in Fig.~\ref{Fig1}(c), XRD studies indicated that the lattice constants of $\rm Bi_{3.875}Co_{0.125}O_4S_3$ and $\rm Bi_{3.875}Ni_{0.125}O_4S_3$ were all reduced relative to those of $\rm Bi_4O_4S_3$, similar to the findings from $\rm Bi_{4-x}Mn_xO_4S_3$ (Fig.~\ref{Fig1}(a)). The $T_{c,\rho}$ and $T_{c,M}$ values only exhibited small variations from those of the parent compound $\rm Bi_4O_4S_3$, as shown in Fig.~\ref{Fig9}(a)-(d). On the other hand, the positive saturation magnetizations in the FC curves for both $\rm Bi_{3.875}Co_{0.125}O_4S_3$ and $\rm Bi_{3.875}Ni_{0.125}O_4S_3$ were smaller than that of $\rm Bi_{3.875}Mn_{0.125}O_4S_3$ (Fig.~\ref{Fig10}), whereas the corresponding ZFC curves remain {\it positive} at all temperatures, in stark contrast to the strong {\it diamagnetism} developed in $\rm Bi_{3.875}Mn_{0.125}O_4S_3$ at $T < 40$ K (Fig.~\ref{Fig2}). Moreover, no enhancement of superconductivity relative to the parent compound $\rm Bi_4O_4S_3$ was found in either $\rm Bi_{4-x}Co_xO_4S_3$ or $\rm Bi_{4-x}Ni_xO_4S_3$ so that anomalies associated with superconductivity in the $M$-vs.-$H$ loops completely vanished at $T >~ T_c = 4.8$ K, as exemplified in Fig.~\ref{Fig6}(a)-(b) for $\rm Bi_{3.875}Co_{0.125}O_4S_3$. 

XPS studies of $\rm Bi_{4-x}Ni_xO_4S_3$ and $\rm Bi_{4-x}Co_xO_4S_3$ (Fig.~\ref{Fig8}(b)-(d)) further revealed that the valences of Ni ions were either 2+ or 3+ and that of Co ions was purely 2+, in contrast to the 3+ and 4+ valences of Mn-ions in $\rm Bi_{4-x}Mn_xO_4S_3$ (Fig.~\ref{Fig8}(a)). These comparisons suggest that the unique 4+ valence of Mn-ions may play an important role in the enhanced superconductivity and ferromagnetism in $\rm Bi_{4-x}Mn_xO_4S_3$ by contributing excess conducting electrons to the superconducting $\rm BiS_2$ layers while retaining localized magnetic moments in the $\rm Bi_2O_2$ spacer layers. The onset of diamagnetic signals at $T <\sim 40$ K in the ZFC magnetization curves of $\rm Bi_{4-x}Mn_xO_4S_3$ (Fig.~\ref{Fig2} and Fig.~\ref{Fig10}) further provides a tantalizing hint for reaching even higher-$T_c$ values in the $\rm BiS_2$-based superconductors. 

\section{Conclusion}

We have demonstrated in this work thermodynamic evidences for enhanced superconductivity and its coexistence with ferromagnetism in Mn-doped layered superconductors, $\rm Bi_{4-x}Mn_xO_4S_3$ ($\rm 0.075 \le x \le 0.15$). Our studies suggest that the robustness of superconductivity against ferromagnetism in these $\rm BiS_2$-based superconductors may be attributed to the layered structure and the significant energy separation between the localized 3$d$ orbitals responsible for magnetism and the Fermi level within the 6$p$ orbitals for superconductivity. In particular, Mn-doping induces an enhancement of $T_c$ from 4.5 K up to $\sim 15$ K, whereas comparable doping of Ni and Co reveals only coexistence of superconductivity and ferromagnetism without discernible enhancement in $T_c$ relative to $\rm Bi_4O_4S_3$. We attribute the unique $T_c$ enhancement in $\rm Bi_{4-x}Mn_xO_4S_3$ to the Mn$^{4+}$/Mn$^{3+}$ mixed valences that contribute excess electrons to the superconducting $\rm BiS_2$ layers, which has been further corroborated by the Hall effect studies. These findings have therefore revealed new pathways to enhancing the $T_c$ of $\rm BiS_2$-based layered superconductors. However, complete elucidation of the microscopic mechanism and the interplay between superconductivity and ferromagnetism in the $\rm BiS_2$-based superconductors still awaits future development of single crystalline materials.

\begin{acknowledgments}
The research at Shanghai University was supported by the Chinese Ministry of Science and Technology (2016YFB0700504), Shanghai Pujiang Program (13PJD015), Science \& Technology commission of Shanghai Municipality (13ZR1415200, 13JC1402400, 11dz1100305), and National Natural Science Foundation of China (NSFC, No. 51372149, 51371111, 51302249, 11204171). The authors acknowledge the technical support by the beamline BL14B1 of Shanghai Synchrotron Radiation Facility for the XPS studies. The research at Caltech was supported by the National Science Foundation under the Institute for Quantum Information and Matter, and by the Moore and Kavli Foundations through the Kavli Nanoscience Institute at Caltech.
\end{acknowledgments}

\end{document}